# Current induced multi-mode propagating spin waves in a spin transfer torque nano-contact with strong perpendicular magnetic anisotropy

S. Morteza Mohseni [1,†], H.F. Yazdi [1], M. Hamdi [1], T. Brächer [2], S. Majid Mohseni [1*]

[1] Faculty of Physics, Shahid Beheshti University, Evin, 19839 Tehran, Iran
[2] Fachbereich Physik and Landesforschungszentrum OPTIMAS, Technische Universität Kaiserslautern, 67663 Kaiserslautern, Germany
[†] Current address: Fachbereich Physik and Landesforschungszentrum OPTIMAS, Technische Universität Kaiserslautern, 67663 Kaiserslautern, Germany

**Abstract**

Current induced spin wave excitations in spin transfer torque nano-contacts are known as a promising way to generate exchange-dominated spin waves at the nano-scale. It has been shown that when these systems are magnetized in the film plane, broken spatial symmetry of the field around the nano-contact induced by the Oersted field opens the possibility for spin wave mode co-existence including a non-linear self-localized spin-wave bullet and a propagating mode. By means of micromagnetic simulations, here we show that in systems with strong perpendicular magnetic anisotropy (PMA) in the free layer, two propagating spin wave modes with different frequency and spatial distribution can be excited simultaneously. Our results indicate that in-plane magnetized spin transfer nano-contacts in PMA materials do not host a solitonic self-localized spin-wave bullet, which is different from previous studies for systems with in plane magnetic anisotropy. This feature renders them interesting for nano-scale magnonic waveguides and crystals since magnon transport can be configured by tuning the applied current.

*Corresponding author's email address: m-mohseni@sbu.ac.ir, majidmohseni@gmail.com

During recent years, there has been an increasing amount of interest in the field of magnonics and magnon-spintronics [1-4] in which magnons are considered as a potential alternative to electrons and photons in wave-based computing systems and information technologies in the near future [5-6]. In this research field, Spin waves, which are the fundamental excitations of a magnetic solid are envisioned as the main candidate for information carriers. They can be generated, manipulated, detected and controlled with different methods including currents [7-8], photons [9-10], phonons [11] etc. One promising way to generate spin waves at the nano-scale is the use of spin polarized currents since the angular momentum of the spin polarized electrons can exert a torque on the magnetic moment of the ferromagnet due to the spin transfer torque (STT) effect [12-13]. The high current density required for STT dependent magnetic excitations can be achieved by injecting a DC current into a nano-contact in spin-vale multilayers where the first layer (fixed layer) works as the polarizer and the second layer (free layer) is the medium in which spin waves are generated and propagated inside [13-15]. These systems are usually known as spin torque oscillators (STOs). One interesting phenomena that occurs in these systems is the mode-coexistence when different spin wave modes can coexist at the same time [16]. This can take place if the spatial symmetry of the system is broken, which indeed can happen, e.g., if the applied current is high enough to create an Oersted field that can compete with the external field and induce an asymmetric field landscape around the contact. In fact, it has been shown previously that, when these systems are magnetized in the plane, at least two different modes can exist, including one non-linear self-localized soliton (spin-wave bullet) and one propagating mode. The localized bullet is formed in the direction where the symmetry of the system is broken while the propagating mode is not affected by the broken symmetry.

On the other hand, spin transfer torque systems with PMA are a subject of interest since they are capable to operate in higher frequencies in comparison to the systems with in-plane magnetic anisotropy mainly due to the contribution of the anisotropy to the effective field [17], and more importantly they can host a non-linear dissipative magnetic droplet soliton [18,19]. In addition, it has been shown that the ellipticity of the magnetization motion and the propagating spin waves in these materials are different which recently drew attentions to study the spin wave propagation in such materials [20].

Here, by employing micromagnetic simulations we show that in in-plane magnetized (by an external field) systems with strong PMA, current induced broken symmetry does not provide the situation for the formation of the solitonic spin wave bullet since the PMA tends to prevent localization of the modes. Instead, two propagating modes are generated simultaneously including the main mode and a mode with lower frequency but with a confined spatial propagation distribution, since the spin wave dispersion around the NC is not homogenous anymore. These findings are interesting from a physical point of view since they are suggesting that PMA influences the non-linearity of such systems when their magnetization is dragged into the plane by an external field, in a way that it prevents the system to go to the negative non-linearity regime. The negative non-linearity results in localization of the spin waves and generation of a spin wave soliton. This can be confirmed by considering the fact that the frequency of the propagating waves is above the frequency of the ferromagnetic resonance (FMR) [21,22]. Indeed, these results differ qualitatively from previously studied systems with in-plane magnetic anisotropy, where analytical calculations, simulations and experiments [22-25] confirmed that a second non-linear spin wave soliton mode becomes favourable for the system in addition to the main stable mode.

Magnetization dynamics induced by the STT effect follows the equation of the magnetization dynamics including an STT term which is described by the well-known Landau-Lifshitz-Gilbert-Slonczewski equation [12]

$$\frac{\partial \hat{m}}{\partial t} = -\gamma \hat{m} \times \mu_0 \vec{H}_{eff} + \alpha \hat{m} \times \frac{\partial \hat{m}}{\partial t} - \gamma \mu_0 M_s \sigma(I) f(\vec{x}) \varepsilon \hat{m} \times (\hat{m} \times \hat{M}) \quad (1)$$

where $\gamma/2\pi = 28 GHz/T$ is the gyromagnetic ratio, $\vec{H}_{eff}$ is the effective field of the free layer including externally applied, exchange, demagnetizing and anisotropy fields, $\alpha$ is the gilbert damping factor, $\vec{m}(x,y)$ is the magnetization in the free layer whereas $\hat{M}$ is the magnetization of the fixed layer, which assumed to be constant in the following. $M_s$ is the saturation magnetization of the free layer and the spin torque defined by the $\sigma(I)$ as the dimensionless spin torque coefficient of the applied current $I$. $\varepsilon$ is the term that defines the spin torque efficiency. $f(\vec{x})$ shows the region where the STT is active which is confined by the size of the nano-contact ($f(\vec{x}) = 1$ inside the NC and zero elsewhere).

The FMR frequency of the system can be calculated based on Kittel's formula including perpendicular magnetic anisotropy [26]

$$\left(\frac{\omega}{\gamma\mu_0}\right)^2 = H_{res}\left(H_{res} + M_s - H_k\right) \quad (2)$$

where $H_{res}$ is the resonance field and $H_k$ is the anisotropy field, respectively.

The investigated structure in this study is modelled using the parameters of the free layer of a nano-contact (NC) current driven pseudo-spin valve (Co/Cu/Co-[Ni/Co]$_4$) where the 4nm Co-[Ni/Co]$_4$ is assumed to be the free layer. The spin polarization is defined via the magnetization angle of the fixed layer, which is found by solving the magnetostatic boundary conditions of the Co layer assuming $M_s = 1350 kA/m$.

Micromagnetic simulations are performed using the GPU-based MuMax3.0 [27] tool, which numerically solves the equation (1) in a medium considered as the free layer of the system with the confined dimensions of 1500*1500*4nm and a constant cell size of 5.85*5.85*4 nm, which is below the exchange length of the system [18]. The material parameters taken into account are in accordance with reference [18] related to the free layer and are as follows: $M_s = 737 kA/m$, $\alpha = 0.02$, $A_{exch} = 12.5 pJ/m$, uniaxial perpendicular magnetic anisotropy of $K_u = 443 kJ/m^3$, spin torque asymmetry $\lambda = 1.3$ and polarization $P = 0.43$. The external field is applied in the plane with a constant amplitude of $1\ T$ and temperature is set to zero in all simulations. Absorbing boundaries are taken into account in order to prevent back reflection of the spin waves and interference effects. The DC current is assumed to pass through an infinite (in z-direction) cylinder-like NC with a diameter of $100\ nm$ in the middle of the simulated geometry with negative polarity.

The simulation is performed in two steps: first, the external field is applied in order to relax the system until it reaches its ground state, and then current is injected into the NC. The z component of magnetization, $M_z(x, y, t)$ in each cell is collected for $20\ ns$ with a sampling time equal to $10\ ps$ and, consequently, a numerical Fast Fourier transformation (FFT) is performed on the collected data to obtain the spin-wave propagation in real space.

In order to clarify the role of the Oersted field, we divide the study into two different situations. For the first step, we present the result of the spin-wave propagation in a system where the Oersted field of the NC is not taken into account, while in the second step the Oersted field contribution is considered and discussed.

## I. Spin-wave propagation without Oersted field

In the first step, we consider a case where the Oersted field of the current in the NC is not incorporated. Figure 1(a) shows the current swept frequency spectra of the magnetic excitations. Increasing the current increases (decreases) the frequency (wavelength) of the excited spin waves constantly, as expected since increasing the current density results in a higher STT based on Eq. (1). This trend is also known as the frequency blue shift [16,21]. The green line which represent the FMR frequency of the free layer calculated based on equ. (2) confirms the propagating nature of the excited waves since the frequency of the excited spin waves are well above the FMR frequency [21].

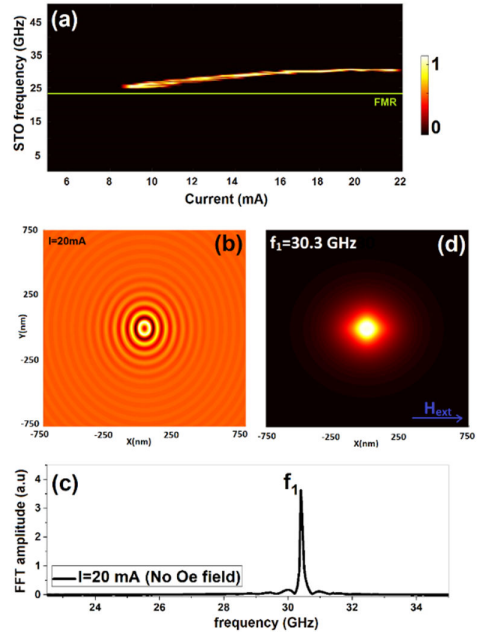

Fig 1. (a) Current swept frequency spectrum of the system at the applied field of $H = 1\ T$ (b) Out of plane component (Mz) of the magnetization oscillation captured at $t = 8\ ns$ after the excitation of the system with the driving current of $I = 20\ mA$ (c) Frequency spectra of the system at $I = 20\ mA$ and (d) Spatial distribution of FFT amplitude of the excited spin-waves at $f_1 = 30.3\ GHz$.

In order to study the propagation and its spatial distribution, a snapshot of the magnetization oscillations at $t = 8\ ns$ after the current injection ($I = 20\ mA$) is shown in fig 1(b) which indicates radial propagation (Slonczewski mode) with a wavelength equal to $60 nm$ that follows the geometry of the NC [12,14,15]. The FFT of the excited waves (fig 1(c)) confirms a single frequency spin wave mode at $f_1 = 30.3\ GHz$ while the spatial distribution of the FFT of this mode which is presented in fig 1(d) exhibits a decay length of $173\ nm$. Increasing the current in this situation where the Oersted field is not considered does not affect the spatial distribution of the waves since the spatial symmetry of the system around the nano contact is not broken by the applied current.

## II. Spin-wave propagation with considering the Oersted field

In this section, we consider the role of the Oersted field induced by the DC current of the NC in all simulations. Fig. 2(a) shows the current swept frequency spectra of the system. Although the frequency of the main mode follows the applied current (blue shift), at $I = 17.7\ mA$ a second mode starts to appear, suggesting that the symmetry of the system is clearly broken by the Oersted field from this point and beyond. The frequency of the second mode decreases with increasing the current (red shift), demonstrating a different nature from the main stable mode [21]. Meanwhile, it should be noted that the frequency of both modes are well above the FMR frequency, indicating their propagating nature [12,21]. As mentioned earlier, in systems with in-plane anisotropy, the frequency of the second mode is below the FMR frequency confirming its solitonic and localized character [16,22,24]. These results are suggesting that the PMA tends to prevent the system to function in a negative non-linearity regime when the magnetization is pulled in-to the plane by an external field even if when the spatial symmetry around the NC is broken.

In order to have a better visualization over the propagating modes, a snapshot of the magnetization oscillation at the applied current $I = 12\ mA$ and $I = 20\ mA$ is presented in fig 2 (b) and 2(c) respectively. Although at $I = 12\ mA$ the propagation of the spin waves shows a radial and symmetric trend, at $I = 20\ mA$, where the two modes appear in the current swept spectrum, the propagation is not symmetric anymore. In particular, the amplitude of the spin waves just above the NC is higher. The FFT of the excited waves for the applied currents $I = 12\ mA$ and $I = 20\ mA$ is shown in fig 2 (d) and 2(e), respectively. The single mode frequency of the spin waves at $I = 12\ mA$ is $27.5\ GHz$. While the frequency of the main mode at $I = 20\ mA$ is $f_1 = 30.4\ GHz$ which is slightly higher than the excited waves without the contribution of the Oersted field (section I), the frequency of the second mode is $f_2 = 27.25\ GHz$ which obviously is lower than $f_1$. It is worth mentioning that the corresponding wavelength of the main mode ($f_1$) is $60nm$, while the wavelength of the second mode ($f_2$) is $90nm$.

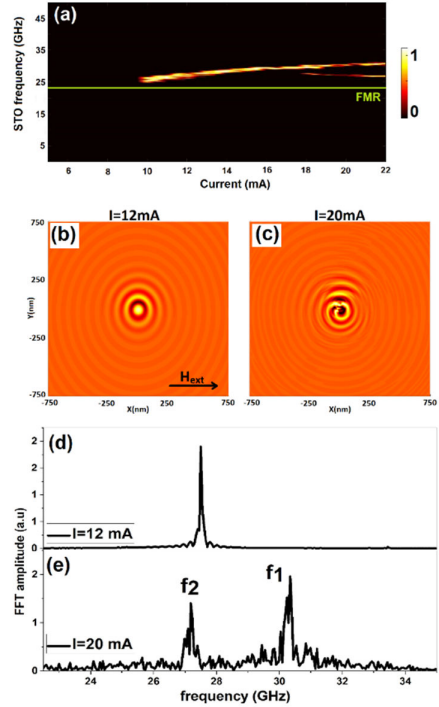

Fig 2. (a) Current swept frequency spectrum of the system at the applied field of $H = 1\ T$ (b) Out of plane component (Mz) of the magnetization oscillation captured at $t = 8\ ns$ after the excitation of the system with the driving current of (b) $I = 20\ mA$ and (c) $I = 12\ mA$, (d) Corresponding Frequency spectra of the system at $I = 12\ mA$ and (e) at $I = 20mA$.

In order to clarify the spatial distribution of the propagating modes, the spatial FFT amplitude of the excited waves in the case of the single mode excitation at $I = 12\ mA$ is presented in fig 3(a) where it indicates a symmetric propagation and a decay length equal to $105nm$.

Fig 3(b) shows the spatial FFT amplitude of the main mode ($f_1$) at the driving current of $I = 20\ mA$. The amplitude of this mode is mainly localized at the bottom of the NC where the Oersted field adds up to the external field. In contrast, the localization of the second mode is mainly above the NC where the Oersted field is oriented against the external field, and thus, tends to reduce it. The amplitude of the Oersted field at the edge of the NC assuming it as an infinite wire with a diameter of $100\ nm$ is equal to $0.08\ T$ based on $h_{Oe} = \mu_0 I/2\pi r$, which is high enough to break the spatial symmetry since it is not negligible with respect to the applied field of $1\ T$. The effective field component parallel to the applied field is shown in fig 3(d) which confirms the broken symmetry around the NC induced by the current. This lower effective field above the NC shifts the dispersion to lower frequencies and this shift confine the spatial distribution of this mode which finally creates a spin-wave beam which exhibit a decay length of $180\ nm$.

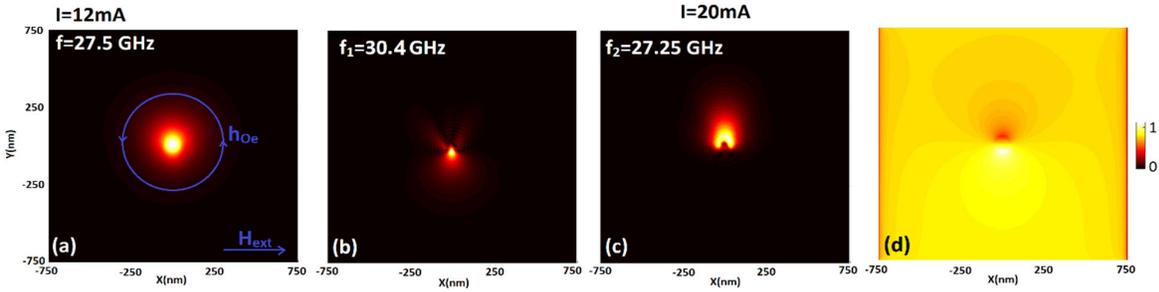

Figure 3. Spatial distribution of the FFT amplitude of (a) the single mode wave at the driving current of $I = 12\ mA$, (b) the main mode at $I = 20\ mA$, (c) the second mode at $I = 20\ mA$, (d) amplitude of the effective field parallel to the external applied field while the NC driven by $I = 20\ mA$.

In conclusion, we have studied the spin-wave excitation and propagation in a system with strong PMA driven by a spin transfer torque nano-contact. Our results show that when such systems are magnetized in the film plane by a strong external field, broken spatial symmetry induced by the Oersted field of the NC provide an opportunity for mode-coexistence which is proven by the excitation of two different propagating spin wave modes at the same time.

The propagating nature of the exited modes opens the question of non-linearity in these systems when they are magnetized in the film plane, since this differs from previously studied systems with in-plane magnetic anisotropy. In such systems, it has been shown that, broken spatial symmetry makes a non-linear solitonic and a propagating mode favourable. In contrast, in an in-plane magnetized STO with strong PMA such as the one studied here, broken spatial symmetry results in two distinct propagating spin-wave modes. While the underlying nature is beyond the scope of this work, the coexistence of these modes is highly interesting for the engineering of nano-scale magnonic elements and waveguides employing current-driven magnetization dynamics.

## Acknowledgments

S.M. Mohseni acknowledges support from Iran Science Elites Federation (ISEF), Iran National Science Foundation (INSF), Iran Nanotechnology Initiative Council (INIC) and Iran's National Elites Foundation (INEF).